\documentclass[twocolumn]{aastex63}
\usepackage[normalem]{ulem}
\usepackage{verbatim}
\usepackage{amssymb}
\usepackage{amsmath}
\usepackage{xspace}
\usepackage{graphicx}
\usepackage{lipsum}
\usepackage{bbm}

\usepackage{amsmath,mathtools,mathrsfs}
\usepackage{hyperref}
\usepackage{afterpage}
\usepackage{booktabs}

\usepackage[encapsulated]{CJK}
\usepackage{ucs}
\usepackage[utf8x]{inputenc}

\submitjournal{ApJ}
\accepted{September 30, 2020}

\newcommand{\V}{{\bf V}}
\newcommand{\D}{{\bf D}}
\newcommand{\G}{{\bf G}}
\newcommand{\CT}{{\mathcal T}}

\newcommand{\CC}{{\mathcal C}}
\newcommand{\eye}{\mathbbm{1}}

\defcitealias{M87_PaperI}{Paper~I}
\defcitealias{M87_PaperII}{Paper~II}
\defcitealias{M87_PaperIII}{Paper~III}
\defcitealias{M87_PaperIV}{Paper~IV}
\defcitealias{M87_PaperV}{Paper~V}
\defcitealias{M87_PaperVI}{Paper~VI}

\shorttitle{Closure Traces}
\shortauthors{Broderick and Pesce}

\graphicspath{{./}{figures/}}

\begin{document}
\title{Closure Traces: Novel Calibration-Insensitive Quantities for Radio Astronomy}

\correspondingauthor{Avery E. Broderick}
\email{abroderick@perimeterinstitute.ca}

\author[0000-0002-3351-760X]{Avery E. Broderick}
\affiliation{ Perimeter Institute for Theoretical Physics, 31 Caroline Street North, Waterloo, ON, N2L 2Y5, Canada}
\affiliation{ Department of Physics and Astronomy, University of Waterloo, 200 University Avenue West, Waterloo, ON, N2L 3G1, Canada}
\affiliation{ Waterloo Centre for Astrophysics, University of Waterloo, Waterloo, ON N2L 3G1 Canada}

\author[0000-0002-5278-9221]{Dominic W. Pesce}
\affiliation{Center for Astrophysics $|$ Harvard \& Smithsonian, 60 Garden Street, Cambridge, MA 02138, USA}
\affiliation{ Black Hole Initiative at Harvard University, 20 Garden Street, Cambridge, MA 02138, USA}

\begin{abstract}
Closure phases and closure amplitudes have proven critical to modern radio interferometry due to their insensitivity to the uncertain station gains.  We present the first set of closure quantities constructed from parallel-hand and cross-hand visibilities that are insensitive to both station gains and to polarimetric leakage.  These complex ``closure traces'' are a natural extension of closure amplitudes and closure phases, are independent of all station-based linear corruptions of the polarized visibilities, and are complete in the sense that they contain all remaining information present in the visibility data.  Products of closure traces on so-called ``conjugate'' quadrangles are sensitive only to structure in the source polarization fraction -- independent of variations in the Stokes $I$ structure -- and thereby provide unambiguous probes of polarization in astronomical sources.
\end{abstract}


\section{Introduction} \label{sec:intro}
Radio antennas are natively sensitive to the polarization of incident radiation, and radio interferometers thus necessarily produce polarized data products.  Astronomical objects that emit in the radio, and in particular the compact sources observable using very long baseline interferometry (VLBI), are also frequently polarized.

In unpolarized radio interferometry, the dominant systematic noise component is typically attributable to complex station-based ``gain'' effects that modulate both the amplitude and the phase of the observed signal \citep{TMS}.  These gain uncertainties are often difficult to calibrate, particularly for VLBI observations at high observing frequencies \citep[e.g.,][hereafter Paper III]{M87_PaperIII}, and they have historically motivated the construction and use of ``closure'' quantities that are unaffected by direction-independent, station-based effects.  This immunity to gain corruptions ensures that both closure phases \citep[introduced by][]{Jennison_1958} and closure amplitudes \citep[introduced by][]{Twiss_1960} are ``robust'' interferometric observables, and it underpins their utility for calibration-independent analyses of radio sources \citep[e.g.,][]{Rogers_1974,Readhead_1980}.

The theoretical foundation for the response of a radio interferometer to polarized radiation was originally derived by \cite{Morris_1964}, and the modern formulation in terms of Jones matrices \citep{Jones_1941} -- the so-called ``radio interferometer measurement equation'' -- was developed in \cite{Hamaker_1996} and further detailed in \cite{Sault_1996}, \cite{HamakerBregman_1996}, and \cite{Hamaker_2000}.  In addition to station gains\footnote{Station gains can be feed-specific, leading to a non-vanishing ``phase-zero differences'' that can dominate the systematic uncertainty of cross-hand quantities.}, polarimetric data products are susceptible to systematic corruption arising from cross-talk between the two separate feeds at each station in the array.  For mixed (``cross-hand'') correlation products -- such as those between the right-hand circular feed of one station and the left-hand circular feed of another -- this polarization leakage  (or ``D terms'') can be the dominant source of systematic calibration uncertainty \citep{RobertsWardleBrown94}.

In this paper, we introduce a new class of closure quantities -- dubbed ``closure traces'' -- that are immune to both gain and leakage corruptions.  Other authors have developed data products that are similarly immune for unpolarized sources \citep{RobertsWardleBrown94} and for point sources \citep{Smirnov_2011}, but to our knowledge the closure traces presented here are the first such quantities to be applicable for radio sources with arbitrary structure and polarization properties.

This paper is organized as follows.  We define the closure traces in \autoref{sec:ClosureTraces}, and in \autoref{sec:Symmetries} we explore their symmetries and degeneracies.  In \autoref{sec:ClosureRelationship} we show how the closure traces reduce to the familiar closure phases and closure amplitudes in the appropriate limits, and in \autoref{sec:IllustrativeApplication} we illustrate the behavior of the closure traces using both synthetic and real datasets.  We summarize and conclude in \autoref{sec:Conclusions}.

\section{Closure Trace Relations} \label{sec:ClosureTraces}
The primary data products in radio interferometry are the complex correlations between the electric fields incident at each telescope in the array, and these correlations provide information about the Fourier transform of the target emission structure via the van Cittert-Zernike theorem \citep{vanCittert_1934,Zernike_1938,TMS}.  The observed polarized visibilities on a baseline between stations $A$ and $B$ are given by the correlation products
\begin{subequations}
\begin{eqnarray}
RR_{AB} & \equiv & \langle E_{R,A} E_{R,B}^* \rangle \\
LL_{AB} & \equiv & \langle E_{L,A} E_{L,B}^* \rangle \\
RL_{AB} & \equiv & \langle E_{R,A} E_{L,B}^* \rangle \\
LR_{AB} & \equiv & \langle E_{L,A} E_{R,B}^* \rangle ,
\end{eqnarray}
\end{subequations}
where $E$ is an electric field, $R$ indicates right-hand circular polarization, $L$ indicates left-hand circular polarization, angular brackets denote a time average, and an asterisk denotes complex conjugation.  We refer to the $RR$ and $LL$ visibilities as ``parallel-hand'' and to the $RL$ and $LR$ visibilities as ``cross-hand'' correlation products.  Throughout this paper we use a circular polarization basis; the same expressions are given for linear or mixed bases in \autoref{app:linfeeds}.

\subsection{Coherency matrix representations of interferometric data}
The combined parallel- and cross-hand visibilities encode the full polarimetric information contained within the Stokes maps of astronomical sources.  Both sets of visibilities are conveniently represented using a coherency matrix formalism.  For a given pair of stations, $A$ and $B$, the associated measurement is
\begin{equation}
\V_{AB}=\left(\begin{array}{cc}
RR_{AB} & RL_{AB}\\
LR_{AB} & LL_{AB}
\end{array}\right).
\end{equation}
The observed $\V_{AB}$ are generally corrupted by station-dependent gains and leakages, which modify the ``true'' $\bar{\V}_{AB}$ via a sequence of linear transformations \citep[see, e.g.,][]{Hamaker_1996},
\begin{equation}
\V_{AB} = \G_A \D_A \bar{\V}_{AB} \D_B^\dagger \G_B^\dagger.
\end{equation}  
Here, $\dagger$ indicates Hermitian conjugation, 
\begin{equation}
\G_{A}=\left(\begin{array}{cc}
G_{R,A} & 0\\
0 & G_{L,A}
\end{array}\right),
\end{equation}
contains the gain terms for station $A$, and
\begin{equation}
\D_{A}=\left(\begin{array}{cc}
1 & D_{R,A}\\
D_{L,A} & 1
\end{array}\right),
\end{equation}  
contains the leakage terms; $\G_{B}$ and $\D_{B}$ are analogous for station $B$.

For circular feeds, the $\V$ are related to the Fourier transforms of the Stokes parameters ($\tilde{I}$, $\tilde{Q}$, $\tilde{U}$, $\tilde{V}$) via 
\begin{equation}
\bar{\V}_{AB} = \left(\begin{array}{cc}
\tilde{I}_{AB}+\tilde{V}_{AB} & \tilde{Q}_{AB}+i\tilde{U}_{AB}\\
\tilde{Q}_{AB}-i\tilde{U}_{AB} & \tilde{I}_{AB}-\tilde{V}_{AB}
\end{array}\right),
\label{eq:VABIQUV}
\end{equation}
and, therefore, the Stokes maps of the source.  
Note that, since the brightness distribution of any Stokes parameter is real-valued and the Fourier transform of any real-valued function is Hermitian, we have
\begin{equation}
\V_{AB} = \V_{BA}^\dagger.
\end{equation}

\subsection{4-station closure traces}
We define a complex-valued closure quantity constructed from the $\V$ defined above and measured on baselines connecting four stations $\{A,B,C,D\}$,
\begin{equation}
  \CT_{ABCD} = \frac{1}{2}{\rm tr}\left( \V_{AB} \V_{CB}^{-1} \V_{CD} \V_{AD}^{-1} \right) .
\label{eq:JCdef}
\end{equation}
These quantities effectively ``close'' via the properties of the trace for non-degenerate $\V$.  That the $\CT$ are independent of the station-specific $\G$ and $\D$ may be proven by repeated use of the identity,
\begin{equation}
    \V_{CB}^{-1}=(\V_{BC}^\dagger)^{-1}
    =\G_{B}^{\dagger -1} \D_{B}^{\dagger -1} \bar{\V}_{CB}^{-1} \D_C^{-1} \G_C^{-1},
\end{equation}
and the cyclic nature of the trace.
In principle, similar closure traces can be generated for any even number of stations.  However, these are not required in general as the above 4-station closure traces are complete (see \autoref{sec:numerology}).

The relationship between visibility and closure trace measurement uncertainties is complicated by the various matrix multiplications and inversions in \autoref{eq:JCdef}.  In the high signal-to-noise limit, and assuming that the uncertainties on real and imaginary components of the individual correlation quantities are independent, the uncertainty in $\CT$ is given by
\begin{equation}
  \begin{aligned}
    \sigma_{\CT}
    =&
    \frac{1}{2}
    \bigg\{
    \sum_{ij} \bigg[
    \left|\left(\V_{CB}^{-1}\V_{CD}\V_{AD}^{-1}\right)_{ij}\right|^2 \sigma_{AB,ji}^2\\
    &+
    \left|\left(\V_{CB}^{-1}\V_{CD}\V_{AD}^{-1}\V_{AB}\V_{CB}^{-1}\right)_{ij}\right|^2 \sigma_{CB,ji}^2\\
    &+
    \left|\left(\V_{AD}^{-1}\V_{AB}\V_{CB}^{-1}\right)_{ij}\right|^2 \sigma_{CD,ji}^2\\
    &+
    \left|\left(\V_{AD}^{-1}\V_{AB}\V_{CB}^{-1}\V_{CD}\V_{AD}^{-1}\right)_{ij}\right|^2 \sigma_{AD,ji}^2
    \bigg]\bigg\}^{1/2},
  \end{aligned}
\label{eq:CTerr}
\end{equation}
where the indices $ij$ extend over the parallel- and cross-hand components of $\V$ and their corresponding uncertainties.  We provide a derivation of the above expression in \autoref{app:error}.

\subsection{Conjugate Closure Trace Products}\label{sec:CCTP}
For any quadrangle $ABCD$, there is a notion of a conjugate quadrangle $ADCB$ for which in the absence of any polarization the closure traces are related via
\begin{equation}
\CT_{ABCD} = \CT_{ADCB}^{-1},
\end{equation}
(see \autoref{eqn:ClosureTraceExpansion}). This relationship motivates the definition of a compound quantity, the conjugate closure trace product, defined by
\begin{equation}
\CC_{ABCD} = \CT_{ABCD}\CT_{ADCB}
\end{equation}
which is identically unity in the absence of polarization.  Correspondingly, deviations from unity in the $\CC_{ABCD}$ are a calibration-independent signature of source polarization.

Associated with the cancellation that makes the $\CC_{ABCD}$ polarization indicators, is the implication that the constituent $\CT$ are correlated.  Where the interferometric polarization fractions are small, i.e., $|\tilde{Q}|,~|\tilde{U}|,~|\tilde{V}|\ll|\tilde{I}|$, this correlation is strong and the $\CC_{ABCD}$ are much better constrained than either $\CT_{ABCD}$ or $\CT_{ADCB}$.  Because this is frequently the situation of interest, it implies that even in the limit of weak source polarization, the $\CC_{ABCD}$ 
can be discriminating indicators of polarization.

\section{Symmetries and degrees of freedom} \label{sec:Symmetries}
A fully-connected array containing four or more stations can be decomposed into a set of non-trivially related quadrangles with all baselines represented among them.  Insofar as the information contained in the visibilities of a single quadrangle can be captured via some composite data product, the information content of the full network of visibilities can necessarily be captured using that data product on a representative set of quadrangles.  Therefore, we now turn to characterizing the symmetries, degeneracies, and completeness of the $\CT$ on a single quadrangle of baselines.

\subsection{General considerations}
Because the $\CT$ are insensitive to gain and leakage effects, any transformation of the $\V$ that can be realized by such effects will manifest as a degeneracy in the $\CT$.  Combining the four degrees of freedom inherent in the two complex gains and two complex leakage terms, the combination $\G_A\D_A$ is any two-dimensional matrix of the form
\begin{equation}
  \G_A\D_A = 
  \left(\begin{array}{cc}
    G_{R,A} & G_{R,A} D_{R,A}\\
    G_{L,A} D_{L,A} & G_{L,A}
  \end{array}\right) .
\end{equation}
Some explicit examples are illuminating.

Choosing for all stations $G_R=G_L=G$, $D_R=0$ and $D_L=0$ corresponds to a rescaling and phase shift of the $\V$ at all stations.  Thus, the $\CT$ are necessarily insensitive to the absolute flux and an over-all phase shift.  Similarly, Choosing for all stations $G_R=G_L=G e^{2\pi i(ux+vy)}$, $D_R=0$ and $D_L=0$ adds a physical shift of the image on the sky.  Because the $\CT$ are defined on quadrangles that close, the additional phase modification necessarily vanishes.  Thus, the $\CT$ suffer from the same sets of degeneracies exhibited by the familiar closure phases and closure amplitudes.

Choosing for all stations $G_R=e^{i\phi/2}$, $G_L=e^{-i\phi/2}$, $D_R=0$ and $D_L=0$ results in 
\begin{equation}
  \G\D = 
  \left(\begin{array}{cc}
    e^{i\phi/2} & 0\\
    0 & e^{-i\phi/2}
  \end{array}\right)
\end{equation}
which induces a rotation of the Stokes sphere about the $V$ axis by $\phi$.  We thus have
\begin{equation}
\V' = \G\D \V (\G\D)^\dagger
=\left(\begin{array}{cc}
\tilde{I}+\tilde{V} & \tilde{Q}'+i\tilde{U}'\\
\tilde{Q}'-i\tilde{U}' & \tilde{I}-\tilde{V}
\end{array}\right),
\end{equation}
where
\begin{equation}
\begin{aligned}
\tilde{Q}' &= \cos\phi \tilde{Q} - \sin\phi \tilde{U}\\
\tilde{U}' &= \sin\phi \tilde{Q} + \cos\phi \tilde{U}.
\end{aligned}
\end{equation}
For circular feeds, a differential gain phase modification can thus rotate the linear polarization's electric-vector position angle (EVPA), implying that the $\CT$ cannot contain any absolute information regarding the EVPA.\footnote{For linear feeds, this corresponds to a rotation in ellipticity, i.e., a rotation about the Stokes $Q$ axis.}

Similarly, choosing for all stations $G_R=\cos(\theta/2)$, $G_L=\cos(\theta/2)$, $D_R=\tan(\theta/2)$ and $D_L=-\tan(\theta/2)$, results in
\begin{equation}
  \G\D = 
  \left(\begin{array}{cc}
    \cos(\theta/2) & \sin(\theta/2)\\
    -\sin(\theta/2) & \cos(\theta/2)
  \end{array}\right)
\end{equation}
which induces a different rotation of the Stokes sphere, this time about the $U$ axis by $\theta$.\footnote{For linear feeds, this corresponds to a rotation in EVPA, i.e., about the Stokes $V$ axis.}  That is,
\begin{equation}
\V'
=\left(\begin{array}{cc}
\tilde{I}+\tilde{V}' & \tilde{Q}'+i\tilde{U}\\
\tilde{Q}'-i\tilde{U} & \tilde{I}-\tilde{V}'
\end{array}\right),
\end{equation}
in which
\begin{equation}
\begin{aligned}
\tilde{Q}' &= \cos\theta \tilde{Q} - \sin\theta \tilde{V}\\
\tilde{V}' &= \sin\theta \tilde{Q} + \cos\theta \tilde{V}.
\end{aligned}
\end{equation}
More generally, the $\CT$ are insensitive to any coherent rotation of the Stokes sphere as seen by all baselines within the quadrangle on which closure trace is defined.
This Stokes orientation degeneracy is the polarimetric analogue of the loss of an absolute phase and absolute flux normalization when using closure phases and closure amplitudes, respectively.  However, we note that the {\em relative} orientation of the Stokes vector as measured on different baselines is preserved.

Insensitivity to arbitrary linear transformation also implies that the $\CT$ can be constructed from arrays containing heterogeneous feeds \citep[see, e.g.,][]{MartiVidal_2016}.  For example, circular feeds are related to linear feeds via a linear transformation given by
\begin{equation}
{\bf Q}_{CL}
=
\frac{1}{\sqrt{2}}
\left(\begin{array}{cc}
1 & 1\\
i & -i
\end{array}\right),
\end{equation}
to which the $\CT$ are invariant \citep[see Equation 5 of][]{MartiVidal_2016}.

\subsection{Closure trace accounting}
\label{sec:numerology}
The closure phases and closure amplitudes are known to contain all of the gain-independent degrees of freedom in Stokes $I$ visibilities \citep[see, e.g.,][]{Blackburn_2020}, and it is natural to ask whether the same property holds for the closure traces with regards to the parallel- and cross-hand visibilities.  The cyclic nature of the trace immediately implies that $\CT$ is symmetric under cyclic permutations of the baselines,
\begin{equation}
\CT_{ABCD}=\CT_{CDAB}.
\end{equation}
In addition, the relationship between baseline reversal and Hermitian conjugation (e.g., $RR_{AB} = RR_{BA}^*$) implies that
\begin{equation}
\CT_{ABCD}^* = \CT_{DCBA}.
\end{equation}
There are thus 6 nonredundant complex $\CT$ out of the 24 that may be constructed from the baselines connecting four independent stations.  With the inclusion of autocorrelation quantities\footnote{We note that such autocorrelations are not typically provided as part of standard VLBI correlator output, and in practice their construction would require more care to avoid systematic biases associated with coherently averaging thermal noise.  Potential mitigation strategies include subsampling in time or frequency.}, e.g. $AA$, there are 4 additional complex $\CT$ that may be constructed.  The remaining $\CT$ are either trivial or may be constructed from these 10.

For example, take the 10 nonredundant complex $\CT$ to be:
$\CT_{ABCD}$, $\CT_{ABDC}$, $\CT_{ACBD}$, $\CT_{ACDB}$, $\CT_{ADBC}$, $\CT_{ADCB}$, $\CT_{ACBA}$, $\CT_{AABC}$, $\CT_{AACD}$, $\CT_{AADB}$, all of which are highlighted in bold below.  The remaining $\CT$ without autocorrelations are given by
\begin{equation}
  \begin{aligned}
    \CT_{BADC}^* &=& \CT_{CDAB}   &=& \CT_{DCBA}^* &=& {\bf \CT_{ABCD}}~\\
    \CT_{BACD}^* &=& \CT_{CDBA}^* &=& \CT_{DCAB}   &=& {\bf \CT_{ABDC}}~\\
    \CT_{BDAC}   &=& \CT_{CADB}^* &=& \CT_{DBCA}^* &=& {\bf \CT_{ACBD}}~\\
    \CT_{BDCA}^* &=& \CT_{CABD}^* &=& \CT_{DBAC}   &=& {\bf \CT_{ACDB}}~\\
    \CT_{BCAD}   &=& \CT_{CBDA}^* &=& \CT_{DACB}^* &=& {\bf \CT_{ADBC}}~\\
    \CT_{BCDA}^* &=& \CT_{CBAD}   &=& \CT_{DABC}^* &=& {\bf \CT_{ADCB}}.
  \end{aligned}
  \label{eq:ABCD_degeneracies}
\end{equation}
The autocorrelation $\CT$ obey the same degeneracies.  For concreteness, consider the $\CT$ generated from baselines between stations in the set $\{A,A,B,C\}$ (i.e., the autocorrelation is measured at station $A$).  In addition, to the above degeneracies, we have
\begin{equation}
  \begin{aligned}
    \CT_{BAAC}   &=& \CT_{CAAB}^* &=& \CT_{ABCA}^* &=& {\bf \CT_{ACBA}} &\\
    \CT_{BCAA}   &=& \CT_{CBAA}^* &=& \CT_{AACB}^* &=& {\bf \CT_{AABC}} &\\
    \CT_{BACA}^* &=& \CT_{CABA}^* &=& \CT_{ACAB}   &=& \CT_{ABAC} &= 1,
  \end{aligned}
  \label{eq:AABC_degeneracies}
\end{equation}
where the $\CT$ on final line are identically unity\footnote{These unit $\CT$ provide a natural set of trivial closure quantities that may be used to assess polarimetric data quality \citepalias[see, e.g.,][]{M87_PaperIII}.}. Similar degeneracies exist for quadrangles constructed with stations $\{A,A,C,D\}$ and $\{A,A,B,D\}$, completing the family of closure traces constructed with autocorrelation measurements at station $A$.  However, there are two additional degeneracies across $AA$ autocorrelation families,
\begin{equation}
  \begin{aligned} 
    \CT_{ADCA} &= {\bf \CT_{AACD}} \frac{\bf \CT_{ADCB}}{\bf \CT_{ABCD}} \frac{\bf \CT_{ACBA}^*}{\bf \CT_{AABC}^*}\\
    \CT_{ABDA} &= {\bf \CT_{AADB}} \frac{\bf \CT_{ADBC}^*}{\bf \CT_{ACBD}^*} \frac{\bf \CT_{ACBA}^*}{\bf \CT_{AABC}^*},
  \end{aligned}
  \label{eq:triangle_degeneracies}
\end{equation}
that are fully specified by quantities in the list of nonredundant $\CT$.  These expressions are derived in \autoref{app:degeneracies}.

In total, there are 26 potential complex numbers to be reconstructed from a fully-connected array containing four stations.  Each of the six baselines provides four complex measurements (the components of $\V$), yielding 24 potential quantities.  The trivial ``baseline'' providing the autocorrelation measurements is subject to additional symmetries that reduce it to two real degrees of freedom ($RR_{AA}$ and $LL_{AA}$, which are both manifestly real) that can be captured as a single complex value, and one complex degree of freedom ($RL_{AA}=LR_{AA}^*$). Contained within these 26 complex values are two complex gains and two complex leakage terms for each station, totaling 16 complex calibration quantities.  The remaining 10 complex degrees of freedom,  matching in number the 10 nonredundant closure traces described in this section, must then characterize the non-calibration (i.e., source) information.  In \autoref{app:completeness} we show that indeed these 10 closure traces are independent.
{\bf Hence, within the $\CT$ is encoded the entirety of the non-calibration information contained in the parallel-hand and cross-hand visibilities.}

\subsection{Fractional source content}
As the number of array stations increases, the fraction of the source content captured by the closure traces tends to unity.  For $N$ stations, the number of complex data quantities (including a single autocorrelation) is
\begin{equation}
N_D = 2 N (N-1) + 2.
\end{equation}
Because the closure traces encode all of the non-calibration information contained in the parallel-hand and cross-hand visibilities, the number of unique $\CT$ is
\begin{equation}
N_{\CT} = N_D - 4 N = 2 N( N-3) + 2. 
\end{equation}
The fraction of the total information available in the complex visibilities that pertains to the source structure and polarization -- i.e., the fraction of the visibility content that may be constrained using the closure traces -- is thus
\begin{equation}
f_{\CT} = \frac{N_{\CT}}{N_D} = \frac{N(N-3)+1}{N(N-1)+1} \approx 1 - \frac{2}{N} ,
\end{equation}
where the final expression is in the limit of large $N$.  Thus, for an array of 5 and 20 stations, $\sim50\%$ and $\sim90\%$ of the source information is captured, respectively.

% The fraction of the total information available in the complex visibilities regarding the source structure and polarization -- i.e., the fraction of the visibility content that may be constrained using the closure traces -- is thus

% The complex visibilities contain all of the source information accessible to the closure traces, but they also contain information about the station calibration quantities.  

\section{Closure trace limits} \label{sec:ClosureRelationship}
\subsection{Point sources} \label{sec:PointSourceLimit}
For point sources, i.e., sources that are significantly unresolved by the baselines comprising a quadrangle, the visibilities are effectively constant and equal to their zero-baseline values.  A direct consequence is that the associated $\CT_{ABCD} = \CT_{AAAA}$ constructed on this quadrangle reduce to unity.  This is similar to the behavior for closure amplitudes and closure phases which reduce to trivial values in this limit.

\subsection{Relationship with other closure quantities} 
The $\CT$ are naturally related to the notion of closure amplitudes and closure phases, to which both reduce in the absence of any intrinsic source polarization and for an appropriate set of baselines.  In this case, $\bar{\V}_{AB}=V_{AB} \eye$, where $V_{AB}=RR_{AB}=LL_{AB}$ is the visibility associated with the Stokes $I$ map.  Then,
\begin{equation}
\CT_{ABCD} = \frac{1}{2}{\rm tr}\left( \frac{V_{AB}V_{CD}}{V_{CB} V_{AD}} \eye \right)
=
\frac{V_{AB}V_{CD}}{V_{CB} V_{AD}}.
\end{equation}
The magnitudes of the above $\CT_{ABCD}$ are the standard closure amplitudes, and the arguments are a form of four-station closure phases.

It is possible to recover standard three-station closure phases using the autocorrelation $\CT$. Including one such autocorrelation in the closure traces permits the construction of
\begin{equation}
\CT_{ACBA} = \frac{B_{CBA}}{|V_{BC}|^2 |V_{AA}|}, 
\end{equation}
where $B_{CBA}$ is the standard bispectrum.  Noting that the denominator is purely real, the standard closure phase is then the complex argument of $\CT_{ACBA}$.

Given the degree of freedom accounting in \autoref{sec:numerology}, it is not surprising that the $\CT$ contain the more traditional closure quantities.  In this sense, the $\CT$ is a super-set of such closures.

\subsection{Weakly-polarized limit} \label{sec:WeakPolLimit}
As noted in \autoref{sec:CCTP}, where the interferometric polarization fractions are small, i.e., $|\tilde{Q}/\tilde{I}|\ll1$, $|\tilde{U}/\tilde{I}|\ll1$, and $|\tilde{V}/\tilde{I}|\ll1$, a new symmetry emerges among the $\CT$.  This is related to the degeneracies implicit in the closure amplitudes, and the fact that these must be realized in the unpolarized limit.  Following the expression of the $\CT$ in terms of the Stokes parameters in \autoref{app:pauli}, to second order in the interferometric polarization fractions, the conjugate closure trace products are
\begin{equation}
\CC_{ABCD}
\approx
1 + q_{ABCD}^2 + u_{ABCD}^2 + v_{ABCD}^2,
\label{eq:CTsymmetry}
\end{equation}
where
\begin{equation}
\begin{aligned}
q_{ABCD}
&=
\frac{\tilde{Q}_{AB}}{\tilde{I}_{AB}}
- \frac{\tilde{Q}_{CB}}{\tilde{I}_{CB}}
+ \frac{\tilde{Q}_{CD}}{\tilde{I}_{CD}}
- \frac{\tilde{Q}_{AD}}{\tilde{I}_{AD}}\\
u_{ABCD}
&=
\frac{\tilde{U}_{AB}}{\tilde{I}_{AB}}
- \frac{\tilde{U}_{CB}}{\tilde{I}_{CB}}
+ \frac{\tilde{U}_{CD}}{\tilde{I}_{CD}}
- \frac{\tilde{U}_{AD}}{\tilde{I}_{AD}}\\
v_{ABCD}
&=
\frac{\tilde{V}_{AB}}{\tilde{I}_{AB}}
- \frac{\tilde{V}_{CB}}{\tilde{I}_{CB}}
+ \frac{\tilde{V}_{CD}}{\tilde{I}_{CD}}
- \frac{\tilde{V}_{AD}}{\tilde{I}_{AD}}.
\end{aligned}
\end{equation}
Where the $q_{ABCD}$, $u_{ABCD}$ and $v_{ABCD}$ vanish identically, this induces an additional approximate symmetry, eliminating 5 of the 10 $\CT$.  When the $\CC_{ABCD}\ne1$, necessarily complex polarization structures are present; the converse is not true: where $\CC_{ABCD}=1$ source polarization may still be present.

Identifying and cataloging all of the conditions under which $\CC_{ABCD}=1$ is beyond the scope if this paper, even if it is generally possible.  Nevertheless, there is a broad class of cases for which this is true.  That is, the $q_{ABCD}$, $u_{ABCD}$, and $v_{ABCD}$ vanish independently when,
\begin{equation}
\begin{aligned}
\tilde{Q} &= m_q(1+2\pi i \vec{\delta}_q\cdot\vec{u})\tilde{I}\\
\tilde{U} &= m_u(1+2\pi i \vec{\delta}_u\cdot\vec{u})\tilde{I}\\
\tilde{V} &= m_v(1+2\pi i \vec{\delta}_v\cdot\vec{u})\tilde{I}
\end{aligned}
\end{equation}
where the $m_q$, $m_u$, $m_v$ are arbitrary real constants and $\vec{\delta}_q$, $\vec{\delta}_u$, $\vec{\delta}_v$, are arbitrary constant vectors (the latter terms vanishing due to the closing of the quadrangle).  When the $\vec{\delta}$ are small, these translate into relationships between the Stokes $Q$, $U$, and $V$ maps and a shifted $I$:
\begin{equation}
\begin{aligned}
Q &= m_q \left( I + \vec{\delta}_q\cdot\vec{\nabla} I \right) 
\approx
m_q I(\vec{x} + \vec{\delta}_q)\\
U &= m_u \left( I + \vec{\delta}_u\cdot\vec{\nabla} I \right) 
\approx
m_u I(\vec{x} + \vec{\delta}_u)\\
V &= m_v \left( I + \vec{\delta}_v\cdot\vec{\nabla} I \right) 
\approx
m_v I(\vec{x} + \vec{\delta}_v),
\end{aligned}
\end{equation}
where the approximations are in the limit of small $|\vec{\delta}|\ll |\vec{u}_{\rm max}|^{-1}$.

Thus, deviations from this additional symmetry require at least that the polarization be non-trivial in the sense that it is not uniform across the source and/or not comprised of uniform components with small shifts.

\subsection{Coincident station limit} \label{sec:TrivialClosures}
For some real-world arrays such as the EHT \citep{M87_PaperII}, the presence of two or more co-located stations in the array provides redundant baselines that can be useful for calibration and error budgeting.  For example, in \citetalias{M87_PaperIII}, appropriately constructed ``trivial'' closure phases and log closure amplitudes -- which for perfectly calibrated data should be zero-valued\footnote{The expectation for zero-valued trivial closure quantities assumes that the co-located stations form a baseline with identically zero length.  For real arrays with a finite station separation, extended structure may result in a violation of this assumption and therefore finite values for the closure quantities.} -- were used to assess the magnitudes of systematic uncertainties remaining in the data after calibration.

Given two co-located stations $A$ and $A'$ and any other two stations $B$ and $C$, we can construct trivial quadrangles of the form $ABA'C$ that we dub ``boomerang quadrangles'' because of their collapsed shape.  In the absence of measurement uncertainty, \autoref{eq:JCdef} predicts that such boomerang quadrangles should have real-valued, unit closure traces so long as $\V_{AB} = \V_{A'B}$ and $\V_{AC} = \V_{A'C}$.  Any deviation of the phase of a trivial closure trace from zero, or of the amplitude of a trivial closure trace from unity, can thus be used to identify the presence of station-independent effects in the data.

Similar trivial $\CT$s exist when three or four co-located stations are available, i.e., $\CT_{AAAB}=1$ or $\CT_{AAAA}=1$.  The latter case also occurs for quadrangles composed of baselines that are sufficiently short that the source is effectively a point source (\autoref{sec:PointSourceLimit}).

\section{Illustrative Application} \label{sec:IllustrativeApplication}
\subsection{Simulated data}

\begin{figure}
\begin{center}
\includegraphics[width=\columnwidth]{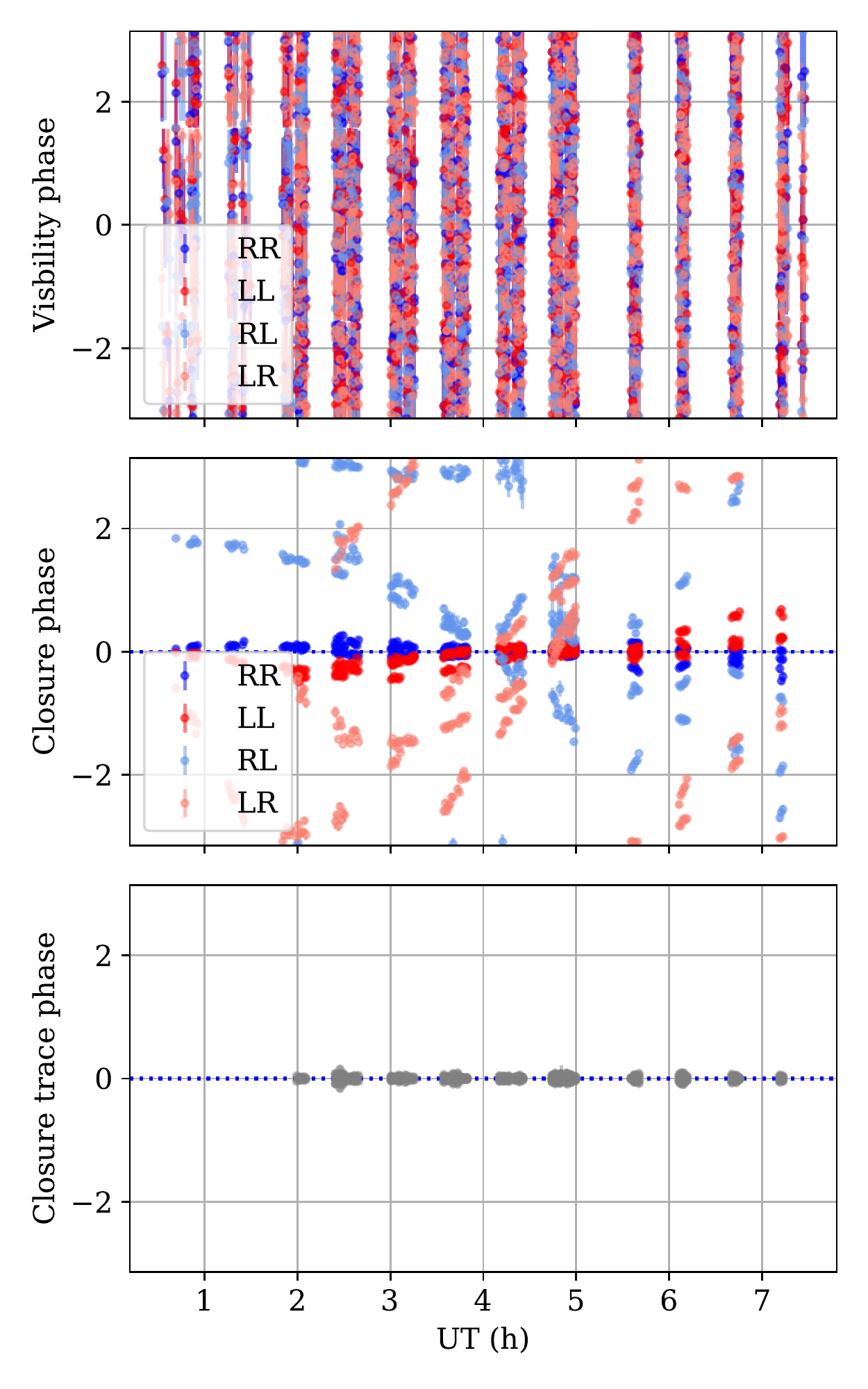}
\end{center}
\caption{Phases of visibilities corresponding to trivial triangles and the trivial boomerang quadrangles (top), trivial closure phases (middle), and trivial boomerang closure trace phases (bottom) for a simulated EHT 2017 observation.  For the top two, these are constructed for the parallel-hand and cross-hand correlation products independently.}
\label{fig:trivial_synth}
\end{figure}

\begin{figure}
\begin{center}
\includegraphics[width=\columnwidth]{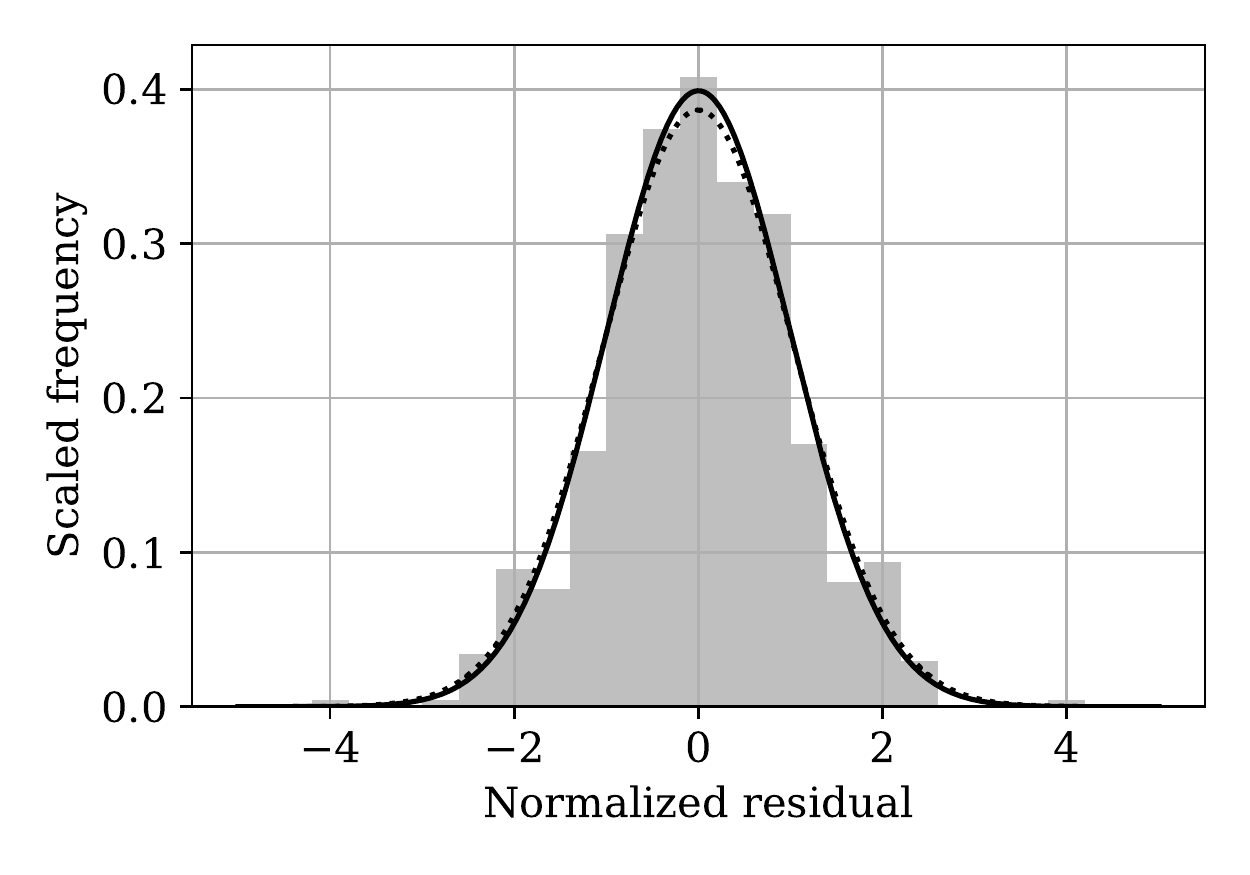}
\end{center}
\caption{Distribution of trivial closure trace phases for a simulated EHT 2017 observation.  For reference a normal distribution with unit variance is shown by the solid black line and a normal distribution with the mean and variance of the residual distribution is shown by the dotted black line.}
\label{fig:trivial_synth_pdf}
\end{figure}

We demonstrate the behavior of the closure traces using a simulated dataset generated using the \texttt{eht-imaging}\footnote{\url{https://github.com/achael/eht-imaging}} library \citep{Chael_2016,Chael_2018} and designed to have the same Fourier coverage as the 2017 April 11 EHT observations \citepalias{M87_PaperIII}.  This dataset includes both station gain and leakage corruptions in addition to realistic baseline-specific thermal errors, with an asymmetric, extended input source emission structure containing nontrivial contributions from all Stokes parameters.  The existence of intra-site baselines between ALMA and APEX presents an opportunity for using closure traces to probe data quality and to assess the magnitude of remaining (i.e., non-closing) systematic errors.

The presence of large atmospheric delays effectively randomizes the phases of individual complex visibilities, shown explicitly in the top panel of \autoref{fig:trivial_synth}.  Closure phases eliminate these large variations. For high signal-to-noise data, and in the absence of polarization leakage, the parallel-hand and cross-hand closure phases constructed on trivial triangles -- ${\rm arg}(B_{ABA'})$, where $A$ and $A'$ are two stations within a single site -- should be distributed normally about $0$ with a width set by the thermal noise.  However, the presence of even modest leakage for a source with significant intrinsic polarization produces large closure phase deviations from zero, as seen in the middle panel of \autoref{fig:trivial_synth}.

Boomerang quadrangles (i.e., those of the form $ABA'C$; see \autoref{sec:TrivialClosures}) are expected to have closure traces that are distributed about unity with a width set by the thermal noise, even in the presence of large gain and leakage errors.  The closure trace phases on these quadrangles should therefore be clustered tightly around 0, as evident in the bottom panel of \autoref{fig:trivial_synth}.  We find that the distribution of the deviations is well described by the thermal noise, as shown explicitly in \autoref{fig:trivial_synth_pdf}.

\begin{figure*}
\centering
\includegraphics[width=\textwidth]{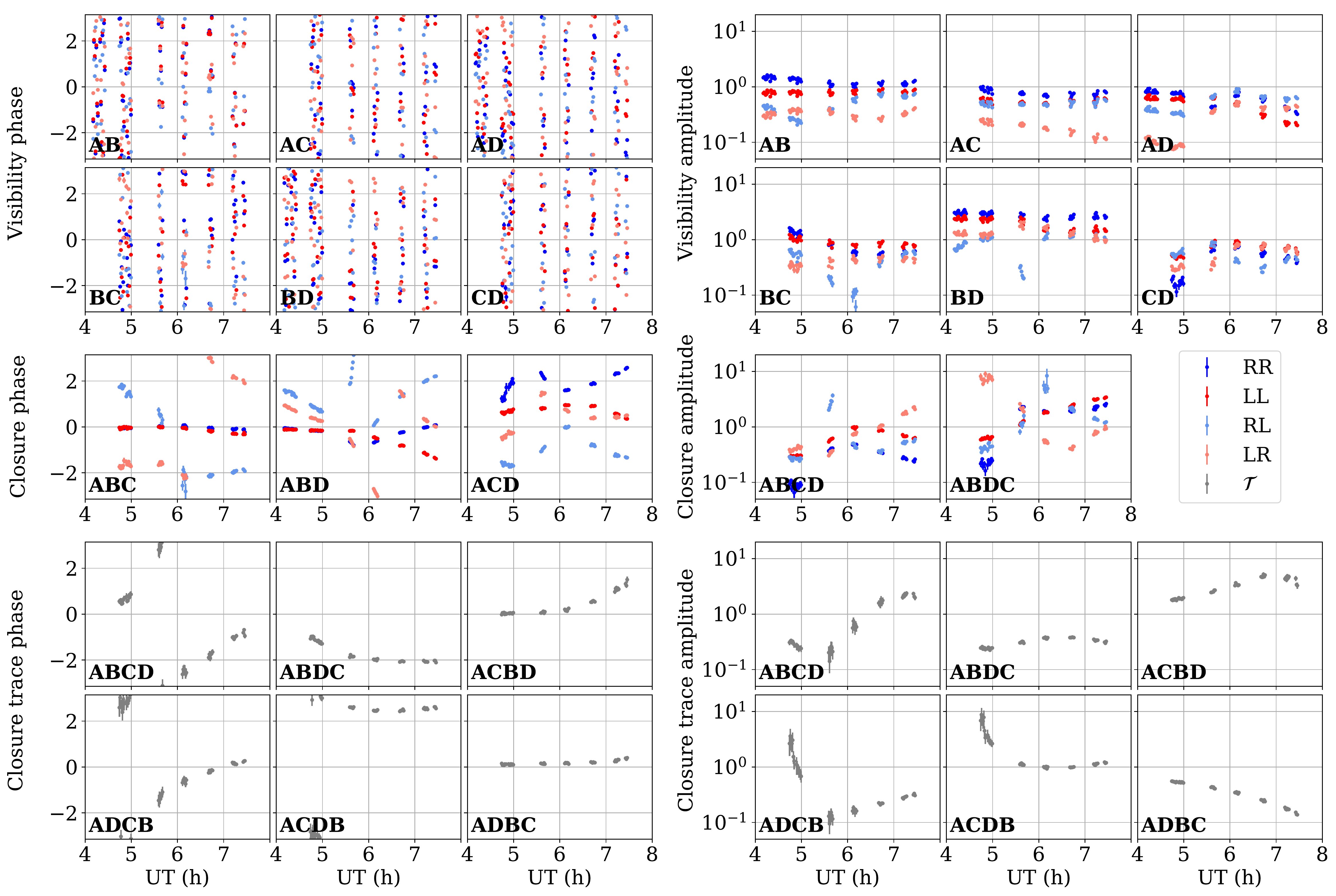}
\caption{Comparison of the visibility, closure, and closure trace phases (left) and amplitudes (right) on the quadrangles composed using four stations for a simulated dataset containing station gain and leakage corruptions.}
\label{fig:nontrivial_synth}
\end{figure*}

An example set of non-trivial baselines, triangles and quadrangles is shown in \autoref{fig:nontrivial_synth}.  Qualitatively, these are similar to the trivial cases.  As before, the visibility phases are randomized by atmospheric delay; the closure phases are more ordered by virtue of their insensitivity to the station-based phase errors though still exhibit large variations due to uncorrected leakage; and the closure traces are considerably more ordered. This is in contrast to the parallel-hand and cross-hand closure phases, which experience more complex trajectories.

\begin{figure*}
\centering
\includegraphics[width=\textwidth]{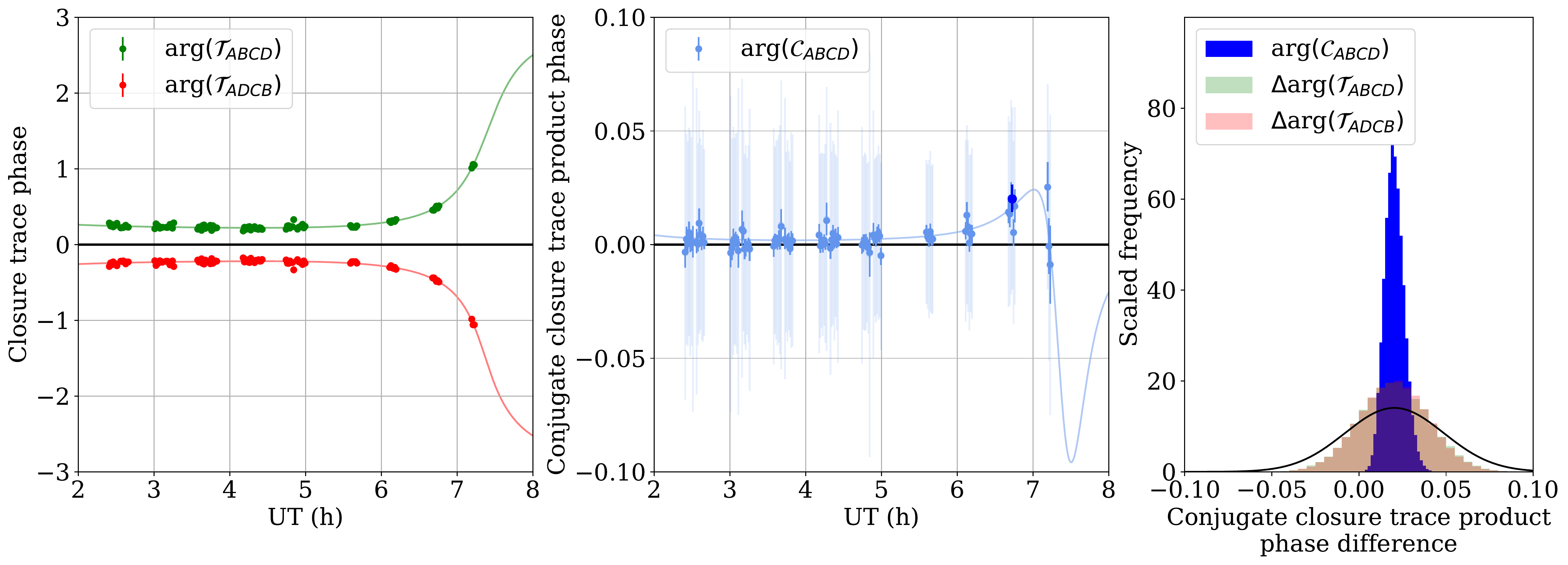}
\caption{Comparison of the closure trace (left) and conjugate closure trace product (center) phases, along with an example set of their thermal distributions (right), on a single quadrangle for a simulated dataset containing station gain and leakage corruptions.  The solid lines passing through the data points in both the left and central panels show the ground-truth behavior of the dataset (i.e., the behavior in the absence of thermal noise).  In the central panel, the light-colored error bars indicate the predicted uncertainty from assuming that the conjugate closure traces are independent, while the dark-colored error bars show the results from Monte Carlo sampling of the visibilities.  The right panel shows the Monte Carlo sampling results in more detail for a single data point, corresponding to the blue highlighted point in the central panel; the solid black curve indicates the expected distribution for the conjugate closure trace product if the conjugate closure traces are assumed to be independent.}
\label{fig:CCTPP_error_budget}
\end{figure*}

\autoref{fig:CCTPP_error_budget} shows the behavior of a selected pair of conjugate closure trace phases and the phase of their corresponding conjugate closure trace product (see \autoref{sec:CCTP}).  The baselines comprising the selected quadrangle have low ($< 10\%$) fractional polarizations, leading to highly symmetric phases for the conjugate closure traces (left panel) and near-zero phase for their product (center panel).  The rightmost panel of \autoref{fig:CCTPP_error_budget} shows the thermal distributions of both conjugate closure trace phases and their corresponding product phase for a single data point, as estimated numerically through Monte Carlo sampling of the constituent visibilities; note that we have shifted the closure trace phase distribution means to coincide with that of the conjugate closure trace product phase distribution.  We can see that the distribution for the conjugate closure trace product phase is considerably narrower than the distributions for either of the individual closure trace phases, and in the case of the selected data point that narrowing makes the difference between a significant detection (i.e. nonzero $\text{arg}(\CC_{ABCD})$) and a complete nondetection (i.e., $\text{arg}(\CC_{ABCD})$ consistent with zero) of polarization structure on this quadrangle.

\subsection{3C 273}
\begin{figure*}
\centering
\includegraphics[width=\textwidth]{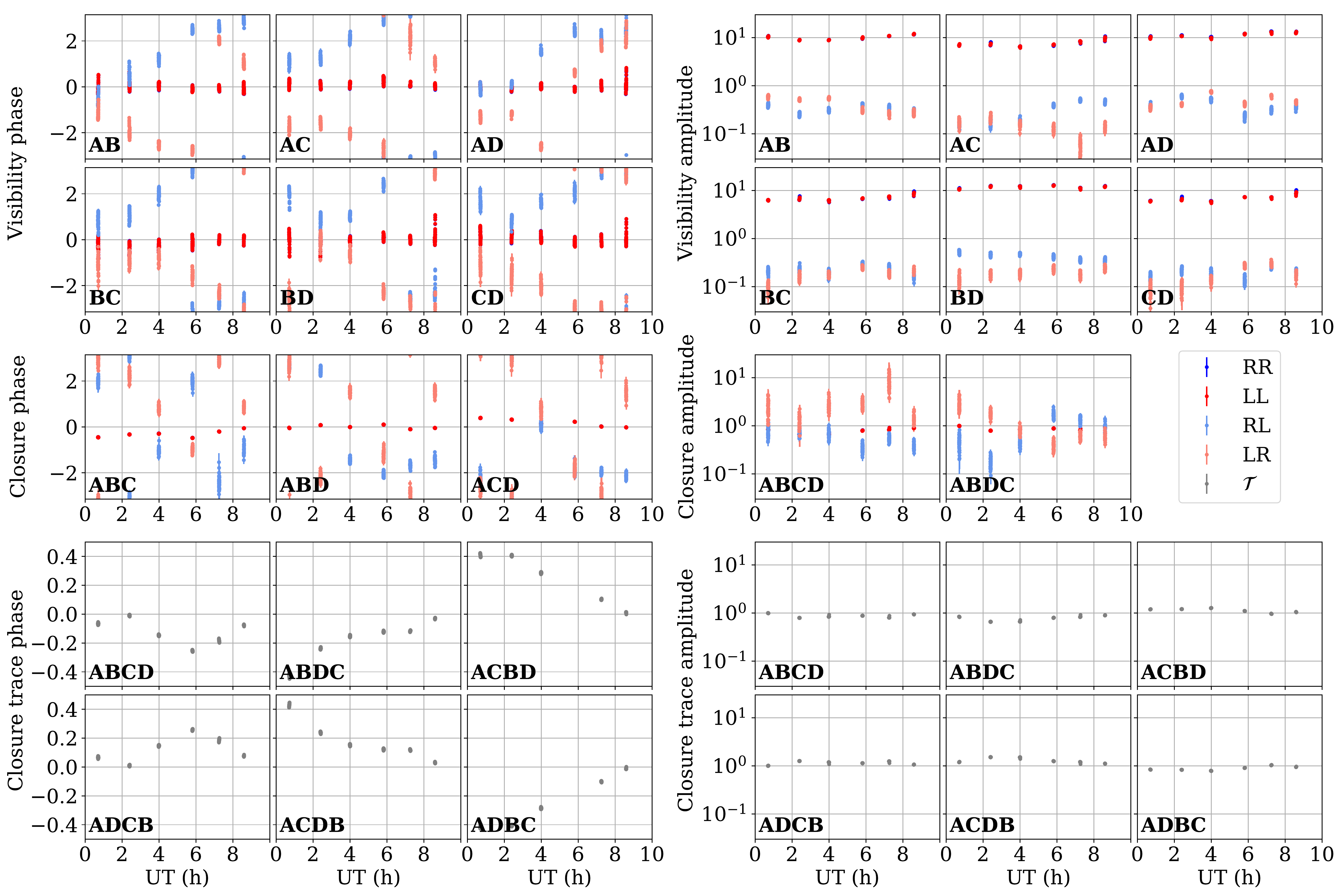}
\caption{Comparison of the visibility, closure, and closure trace phases (left) and amplitudes (right) on the quadrangles composed using the LA, KP, NL, and FD stations for the MOJAVE 3C 273 dataset.}\label{fig:3C273}
\end{figure*}

\begin{figure*}
\centering
\includegraphics[width=\textwidth]{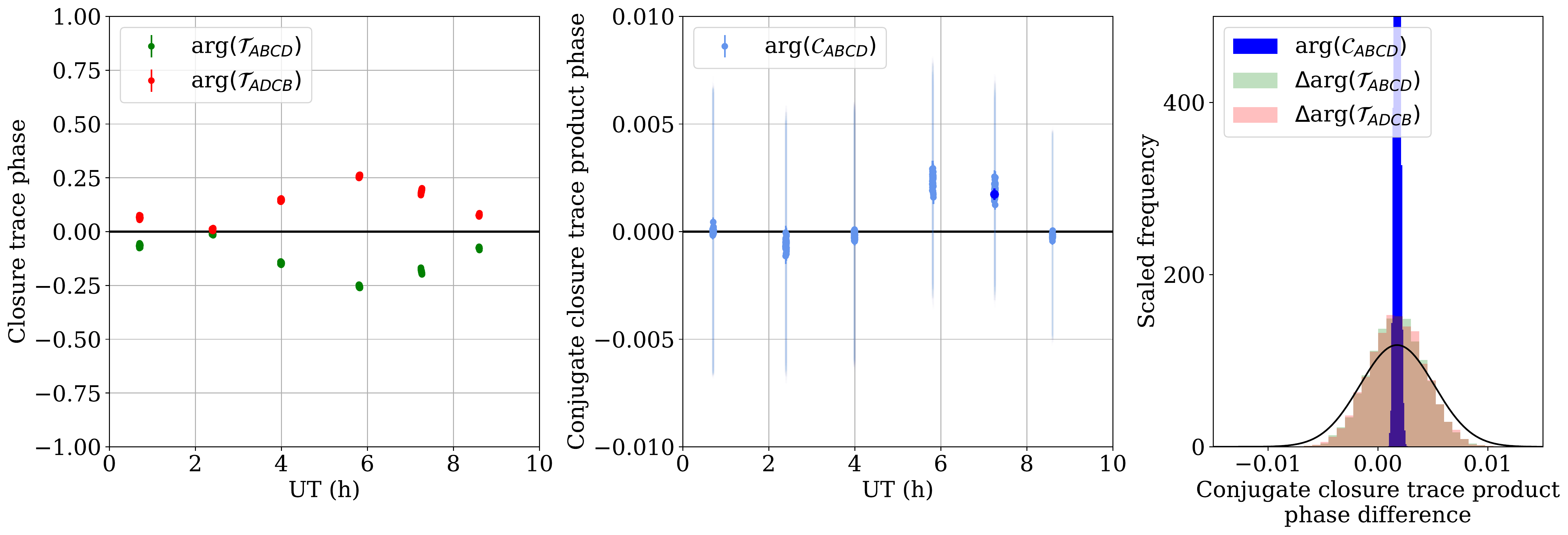}
\caption{Same as \autoref{fig:CCTPP_error_budget}, but using the LA-KP-NL-FD quadrangle from the MOJAVE 3C 273 dataset.
}
\label{fig:CCTPP_error_budget_3C273}
\end{figure*}

We further demonstrate the closure traces on the publicly available\footnote{\url{https://www.physics.purdue.edu/MOJAVE/sourcepages/1226+023.shtml}} MOJAVE 3C 273 data set \citep{Lister_2018}.  3C 273 is a powerful blazar that is very radio bright, well resolved by the VLBA at 15~GHz, highly polarized, and exhibits complex polarization structures.  It thus provides a number of opportunities for non-trivial closure quantities.  We use the data as provided, with no additional calibration or averaging applied; for details regarding the data reduction procedures, see \cite{Lister_2009,Lister_2018}.

Because the $\CT$ are specifically independent of the particulars of the station gain and leakage calibration, we construct the $\CT$ from the uncalibrated data sets for the May 1, 2020 observations, shown in \autoref{fig:3C273}.  Despite the lack of calibration, the observed complex visibilities are much better behaved than those in the high-frequency simulated data set as a consequence of the reduced atmospheric phase delays at long wavelengths.  Nevertheless, the cross-hand visibilities and closure phases fluctuate within scans by considerably more than the inferred thermal noise.  In comparison, all of the closure traces appear well constrained.

There is a large degree of obvious symmetry between pairs of $\CT$, anticipated from \autoref{sec:ClosureRelationship} and arising from the modest polarization fractions throughout the image -- the asymmetries between the conjugate $\CT$ are quadratic in the polarization fraction (see, e.g., \autoref{eq:CTsymmetry}).  MOJAVE images of 3C 273 indicate typical polarization fractions of order 10-20\%, and thus deviations from symmetry of order 1-4\% are expected.  This low polarization is also responsible for the similarities between the parallel-hand visibilities and closure phases, indicated by the apparent absence of dark-blue $RR$ points that are otherwise obscured by the dark-red $LL$ points in the top three rows of \autoref{fig:3C273}.

\autoref{fig:CCTPP_error_budget_3C273} is analogous to \autoref{fig:CCTPP_error_budget} and shows the conjugate closure trace product on the quadrangle LA-KP-NL-FD for the 3C 273 dataset.  When the uncertainties are propagated in a manner that accounts for the correlations between the constituent closure traces, the value of this conjugate closure phase product deviates significantly from zero. The calibration-independent nature of the closure traces makes this behavior an unambiguous indication of the presence of nontrivial fractional polarization structure in 3C 273.

\section{Conclusions} \label{sec:Conclusions}
A class of polarization closure quantities may be constructed on quadrangles over which all parallel-hand and cross-hand visibilities have been measured.  These ``closure traces'' have the following properties:
\begin{enumerate}
\item They are independent of any linear corruption terms that leave the coherency matrix invertible.  These include, though are more general than, the standard complex station gains and leakage terms (i.e., D terms).
\item They are complete in that they encode all remaining source information.
\item They are a superset of the traditional closure amplitudes and closure phases, to which they reduce in the absence of source polarization.
\end{enumerate}

The closure traces are subject to a variety of degeneracies inherent in the uncertain calibration.  These degeneracies include those to which the closure amplitudes and phases are subject (i.e., absolute flux calibration, phase calibration, source position on the sky), as well as arbitrary rotations of the Stokes sphere.  This last degeneracy has the benefit of implying that the closure traces may be constructed directly from arrays with heterogeneous feed geometries\footnote{For example, the EHT in 2017, for which ALMA recorded linear feeds while the remaining stations recorded circular feeds.}.

A number of immediate applications exist for closure traces as a consequence of their independence from station-based calibration effects.  These include:
\begin{itemize}
\item Closure traces computed on trivial quadrangles permit a quantitative assessment of data quality and consistency, similar to prior efforts that employ closure phases \citepalias{M87_PaperIII}.
\item Closure traces computed on nontrivial quadrangles provide a direct probe of source structure and its evolution across subsequent observations.
\item Conjugate closure trace products directly probe structure in the source polarization fraction, independent of variations in the Stokes $I$ map.  These quantities thus provide unambiguous signatures of polarization in astronomical radio sources.
\end{itemize}
We have demonstrated the existence and performance of the closure traces on simulated and real VLBI data sets.

In practice, the use of closure traces differs from the more traditional methods of polarimetric and gain calibration via the assumption of priors.  Typically, leakage terms are expected to be constant over individual observations, if not much longer.  As a result, multiple observing epochs may be combined to produce leakage estimates, reducing the number of calibration degrees of freedom significantly relative to those presumed by the closure traces themselves.  Moreover, often the a priori gain amplitude and leakage calibrations are expected to be accurate, removing the possibility of the large deviations permitted by the closure traces.

Nevertheless, the increased independence of the closure traces presents a novel method for assessing standard gain and leakage calibration schemes, relaxing assumptions about the magnitude and stability of corrupting effects.  As such, they provide a powerful new tool with which to probe polarized VLBI observations.

\acknowledgments
The authors would like to thank Ivan Mart\'i-Vidal, Daniel Palumbo, Lindy Blackburn, Michael D.~Johnson, Jos\'e Gomez, and Sheperd S.~Doeleman for helpful discussions.
We further thank Jos\'e Gomez for making available the uncalibrated May 1, 2020, 15GHz data for 3C 273.
This work was supported in part by Perimeter Institute for Theoretical Physics.  Research at Perimeter Institute is supported by the Government of Canada through the Department of Innovation, Science and Economic Development Canada and by the Province of Ontario through the Ministry of Economic Development, Job Creation and Trade.
A.E.B. thanks the Delaney Family for their generous financial support via the Delaney Family John A. Wheeler Chair at Perimeter Institute.
A.E.B. receives additional financial support from the Natural Sciences and Engineering Research Council of Canada through a Discovery Grant.
D.W.P. is supported in part by the Black Hole Initiative at Harvard University, which is funded by grants from the John Templeton Foundation and the Gordon and Betty Moore Foundation to Harvard University.
This research has made use of data from the MOJAVE database that is maintained by the MOJAVE team \citep{Lister_2018}.

\software{\texttt{eht-imaging} \citep{Chael_2016,Chael_2018}}, Mathematica \citep{Mathematica}

\clearpage
\appendix

\section{Closure Traces for Linear and Mixed Feeds} \label{app:linfeeds}
When the feeds are linear, for a given pair of stations, $A$ and $B$, the measurements are,
\begin{equation}
\V_{AB} = \left(\begin{array}{cc}
 XX_{AB} & XY_{AB} \\
 YX_{AB} & XX_{AB}
\end{array}\right),
\end{equation}
where $X$ and $Y$ correspond to the two orthogonal feeds.  This is related to the Fourier transforms of the Stokes parameters via
\begin{equation}
\V_{AB} = \left(\begin{array}{cc}
\tilde{I}_{AB}+\tilde{Q}_{AB} & \tilde{U}_{AB}+i\tilde{V}_{AB}\\
\tilde{U}_{AB}-i\tilde{V}_{AB} & \tilde{I}_{AB}-\tilde{Q}_{AB}
\end{array}\right).
\end{equation}
Note that these are related to the quantities in \autoref{eq:VABIQUV} via a rotation of the Stokes sphere that moves $Q\rightarrow U$, $U\rightarrow V$, and $V\rightarrow Q$.  As a consequence, the $\CT$ are unchanged.

A similar conclusion applies for mixed feeds, and indeed any feed geometry that captures linearly independent polarization modes.

\section{Closure Trace Error Estimates} \label{app:error}
We begin by noting that the derivative of $\V^{-1}$ with respect to the elements of $\V$ is,
\begin{equation}
  \begin{aligned}
    \left[\frac{\partial \V^{-1}}{\partial V_{ij}}\right]_{mn}
    &=
    -\left[\V^{-1} \frac{\partial\V}{\partial V_{ij}} \V^{-1}\right]_{mn}\\
    &=
    -\sum_{kl} V_{nk}^{-1} \delta_{ik} \delta_{jl} V_{lm}^{-1}\\
    &=
    -V_{ni}^{-1} V_{jm}^{-1}.
  \end{aligned}
\end{equation}
We further note that for any square two-dimensional matrix ${\bf M}$, the partial derivatives with respect to the elements of $\V$ of ${\rm tr}(\V {\bf M})$ are given by
\begin{equation}
    \frac{\partial}{\partial V_{ij}} \frac{1}{2}{\rm tr}\left( \V {\bf M} \right) 
    = 
    \frac{1}{2} \frac{\partial}{\partial V_{ij}} \sum_{m,n} V_{mn} M_{nm}
    =
    \frac{1}{2} \sum_{m,n} \delta_{mi}\delta_{nj} M_{nm}
    =
    \frac{1}{2} M_{ji}.
\end{equation} 
Therefore, after utilizing the definition and cyclic nature of the trace,
\begin{equation}
  \begin{aligned}
    \frac{\partial\CT_{ABCD}}{\partial V_{AB,ij}}
    &=
    \frac{1}{2} \left(\V_{CB}^{-1} \V_{CD} \V_{DA}^{-1} \right)_{ji}\\
    \frac{\partial\CT_{ABCD}}{\partial V_{CD,ij}}
    &=
    \frac{1}{2} \left(\V_{DA}^{-1} \V_{AB} \V_{CB}^{-1} \right)_{ji}\\
    \frac{\partial\CT_{ABCD}}{\partial V_{CB,ij}}
    &=
    \frac{1}{2} \left(\V_{CB}^{-1} \V_{CD} \V_{DA}^{-1} \V_{AB} \V_{CB}^{-1} \right)_{ji}\\
    \frac{\partial\CT_{ABCD}}{\partial V_{DA,ij}}
    &=
    \frac{1}{2} \left(\V_{DA}^{-1} \V_{AB} \V_{CB}^{-1} \V_{CD} \V_{DA}^{-1} \right)_{ji}
  \end{aligned}
\end{equation}
These may be combined in the normal way, under the assumption that the components of each $\V$ are independent from each other, to produce the uncertainty estimate in \autoref{eq:CTerr}.

\section{Degeneracies}\label{app:degeneracies}
\subsection{Traces, Determinants, and Uniqueness} 
In two dimensions there is straightforward relationship between the determinants and traces of invertible matrices.  Given any invertible $2\times2$ matrix ${\bf M}$, we can diagonalize it, giving eigenvalues $m_1$ and $m_2$.  In terms of these, the trace and determinant are given by
\begin{equation}
{\rm tr}({\bf M}) = m_1 + m_2
~~\text{and}~~
\|{\bf M}\| = m_1 m_2.
\end{equation}
Similarly,
\begin{equation}
{\rm tr}({\bf M}^{-1}) = \frac{1}{m_1} + \frac{1}{m_2}
~~\text{and}~~
\|{\bf M}^{-1}\| = \frac{1}{m_1 m_2}.
\end{equation}
Note that two things are immediately apparent:
\begin{enumerate}
\item The two traces are linearly independent generally, implying that it is possible to reconstruct $m_1$ and $m_2$ uniquely from ${\rm tr}({\bf M})$ and ${\rm tr}({\bf M}^{-1})$. 
\item The two determinants are not independent (yielding only a single combination of $m_1$ and $m_2$).
\end{enumerate}
Because both eigenvalues may be reconstructed from the traces, this implies that the determinant may be as well.  Indeed, the ratio of the two traces gives
\begin{equation}
\frac{{\rm tr}({\bf M})}{{\rm tr}({\bf M}^{-1})} = \|{\bf M}\|.
\end{equation}
Often manipulations with determinants are more convenient despite the greater information content of the traces.

\subsection{Triangle Trace Degeneracies}
We now prove the relations in \autoref{eq:triangle_degeneracies}.  We begin by noting that
\begin{equation}
\|\V_{AD}\V_{CD}^{-1}\V_{CA}\V_{AA}^{-1}\|
=
\|\V_{AD}\V_{CD}^{-1}\V_{CB}\V_{AB}^{-1}\|
\|\V_{AB}\V_{CB}^{-1}\V_{CA}\V_{AA}^{-1}\|.
\end{equation}
Using the relationships between traces and the determinants, and simplifying, we obtain
\begin{equation}
\frac{\CT_{ADCA}}{\CT_{AACD}}
=
\frac{\CT_{ADCB}}{\CT_{ABCD}}
\frac{\CT_{ABCA}}{\CT_{AACB}},
\end{equation}
which may be rearranged to solve for $\CT_{ADCA}$ in terms of remaining quantities.  Applying the known degeneracies of the $\CT$ yields the desired relation.

Similarly, to prove the second relation, we start with
\begin{equation}
\|\V_{AB}\V_{DB}^{-1}\V_{DA}\V_{AA}^{-1}\|
=
\|\V_{CB}\V_{DB}^{-1}\V_{DA}\V_{CA}^{-1}\|
\|\V_{AB}\V_{CB}^{-1}\V_{CA}\V_{AA}^{-1}\|
\end{equation}
Again, rewriting this in terms of the traces,
\begin{equation}
\frac{\CT_{ABDA}}{\CT_{AADB}}
=
\frac{\CT_{CBDA}}{\CT_{CADB}}
\frac{\CT_{ABCA}}{\CT_{AACB}},
\end{equation}
which may be re-expressed as in \autoref{eq:triangle_degeneracies} after applying the degeneracies in Equations \ref{eq:ABCD_degeneracies} and \ref{eq:AABC_degeneracies}.

\section{Completeness via continuous symmetries}\label{app:completeness}
We may directly demonstrate the completeness of the $\CT$ via the rank of the Jacobian of their definitions relative to the components of the visibilities.  That is, the matrix
\begin{equation}
J_{q,bij} = \frac{\partial \CT_q}{\partial V_{b,ij}}
\end{equation}
where $q$ runs over the 10 quadrangle combinations ($ABCD$, $ABDC$, $ACBD$, $ACDB$, $ADBC$, $ADCB$, $ACBA$, $AABC$, $AACD$, $AADB$), $b$ runs over the 7 baseline combinations ($AA$, $AB$, $AC$, $AD$, $BC$, $BD$, $CD$), $ij$ runs over the 4 components of each $\V$ ($RR$, $LL$, $RL$, $LR$).  This is further subdivided by the real and imaginary components of each quantity, yielding a $20\times52$ dimensional matrix (note that the symmetries of the $\V_{AA}$ imply that only 4 numbers are required to specify the $V_{AA,ij}$).

That the matrix is not square is an indication of the fact that the set of $\CT$ being considered does not encompass the calibration information, which makes up the remaining 32 dimensions (real and imaginary parts of 16 complex quantities).  

The rank of this matrix was evaluated for arbitrary $V_{b,ij}$ via Mathematica, finding that indeed it is 20.  That is, the 20 combinations of the $V_{b,ij}$ encoded in the real and imaginary components of the 10 $\CT$ are indeed linearly independent, and thus unique.

\section{Expression of Closure Traces in Stokes Parameters}\label{app:pauli}
The coherency matrix can be rewritten in terms of the Pauli matrices \citep{Hamaker_2000},
\begin{equation}
\V = \tilde{I}\left(\eye + s^a \sigma_a\right),
\end{equation}
where $s^a=(\tilde{Q}/\tilde{I},\tilde{U}/\tilde{I},\tilde{V}/\tilde{I})$, $\sigma_a=(\sigma_x,\sigma_y,\sigma_z)$ are the standard Pauli matrices, and summation is implied across repeated up/down indices (i.e., $s^a\sigma_a = s^1\sigma_1 + s^2\sigma_2 + s^3\sigma_3$).  It is straightforward to verify that
\begin{equation}
\V^{-1} =
\frac{\eye - s^a\sigma_a}{\tilde{I}(1-s^b s_b)},
\end{equation}
using $\sigma_a\sigma_b = \delta_{ab}\eye + i \epsilon^c_{~ab} \sigma_c$, where $\epsilon_{abc}$ is the antisymmetric tensor (i.e., the Levi-Civita tensor).  These may be employed to simplify products of coherency matrices, e.g.,
\begin{equation}
\V_{AB} \V_{CB}^{-1}
=
\left(\eye+s_{AB}^a\sigma_a\right)
\frac{\left(\eye-s_{CB}^b\sigma_b\right)}{\left(1-s_{CB}^c s_{CB,c}\right)}
=
\frac{1-s_{AB}^a s_{CB,a}}{1-s_{CB}^d s_{CB,d}} \eye
+
\frac{s_{AB}^c - s_{CB}^c - i\epsilon^c_{~ab} s_{AB}^a s_{CB}^b}{1-s_{CB}^d s_{CB,d}}
\sigma_c.
\end{equation}
Finally, note that by virtue of the linear nature of the trace,
\begin{equation}
{\rm tr}(\V) 
= \tilde{I} {\rm tr}(\eye) + s^a {\rm tr}(\sigma_a)
= 2 \tilde{I},
\end{equation}
where we have used the fact that the Pauli matrices are traceless, i.e., ${\rm tr}(\sigma_a)=0$.  With repeated use of these properties we can compute the closure traces directly in terms of the Stokes parameters:
\begin{equation}
\begin{aligned}
\CT_{ABCD}
&=
\frac{\tilde{I}_{AB}\tilde{I}_{CD}}{\tilde{I}_{AD}\tilde{I}_{CB}}
\bigg[
1 
- 
\left( s_{AB}^a s_{CB,a}
+ s_{CD}^a s_{AD,a}
- s_{AB}^a s_{CD,a}
+ s_{BC}^a s_{CD,a}
+ s_{AB}^a s_{AD,a}
- s_{BC}^a s_{AD,a}
\right)\\
&\qquad\qquad\qquad~
+ i \epsilon_{abc} \left(
  s_{AB}^a s_{CB}^b s_{CD}^c
- s_{AB}^a s_{CB}^b s_{AD}^c
+ s_{AB}^a s_{CD}^b s_{AD}^c
- s_{CB}^a s_{CD}^b s_{AD}^c \right)\\
&\qquad\qquad\qquad~
+ \left( 
s_{AB}^a s_{CB,a} s_{CD}^b s_{AD,b}
- s_{AB}^a s_{CD,a} s_{CB}^b s_{AD,b}
+ s_{AB}^a s_{AD,a} s_{CB}^b s_{CD,b}
\right)
\bigg]\\
&\qquad\qquad\qquad\bigg/\bigg[ 
\left(
    1-s_{CB}^d s_{CB,d} 
\right)\left(
    1-s_{AD}^e s_{AD,e}
\right)\bigg],
\end{aligned} \label{eqn:ClosureTraceExpansion}
\end{equation}
where terms are grouped by order in $s^a$ in the numerator.  Note that there are no linear-order polarization terms in the closure traces.  Because \autoref{eqn:ClosureTraceExpansion} is constructed solely by inner products in the Stokes space with other Stokes parameters and the Levi-Civita tensor, it is also manifestly invariant to rotations of Stokes sphere that coherently act on all of the baselines involved.

Expanding the closure traces to lowest order in the polarization quantities gives
\begin{equation}
\begin{aligned}
\CT_{ABCD} &\approx
\frac{\tilde{I}_{AB}\tilde{I}_{CD}}{\tilde{I}_{AD}\tilde{I}_{CB}}
\left[
1 
+ s_{CB}^a s_{CB,a}
+ s_{AD}^a s_{AD,a}
- s_{AB}^a s_{CB,a}
- s_{CD}^a s_{AD,a} 
\right.\\
&\qquad\qquad\qquad
\left.
+ s_{AB}^a s_{CD,a}
- s_{CB}^a s_{CD,a}
- s_{AB}^a s_{AD,a}
+ s_{CB}^a s_{AD,a} \right].
\end{aligned}
\end{equation}
Therefore, to lowest order, the conjugate closure trace product, corresponding to the product of inverse-pairs, is
\begin{equation}
\CC_{ABCD}
=
1 
+ 
s_{ABCD}^a s_{ABCD,a}
\quad\text{where}\quad
s_{ABCD}^a = s_{AB}^a - s_{CB}^a + s_{CD}^a - s_{AD}^a.
\end{equation}

\bibliography{references}{}
\bibliographystyle{aasjournal}

\end{document}